\documentclass{mem}
\usepackage{natbib}\usepackage{txfonts}\usepackage{balance}
\usepackage{graphicx}
\usepackage{epstopdf}
\usepackage[a4paper]{hyperref}
\newcommand{\cobold}{\mbox{CO$^\mathrm{\,5}$BOLD }}
\begin{document}
\title{The CO5BOLD Analysis Tool} 
   \subtitle{}

\author{S. \, Wedemeyer}

\offprints{S. Wedemeyer}

\institute{
Institute of Theoretical Astrophysics, University of Oslo,
  P.O. Box 1029 Blindern, N-0315 Oslo, Norway
\email{sven.wedemeyer-bohm@astro.uio.no}
}

\authorrunning{Wedemeyer}

\titlerunning{The CO5BOLD Analysis Tool}

\abstract{The interactive IDL-based CO5BOLD Analysis Tool (CAT) was developed 
to facilitate an easy and quick analysis of numerical simulation data 
produced with the 2D/3D radiation magnetohydrodynamics code CO5BOLD. 
The basic mode of operation is the display and analysis of cross-sections 
through a model either as 2D slices or 1D graphs. 
A wide range of physical quantities can be selected.
Further features include the export of models into VAPOR format or the output 
of images and animations. 
A short overview including scientific analysis examples is given. 

}
\maketitle

\section{Introduction}

\begin{figure}[bp!]
\centering
\resizebox{3.7cm}{!}{\includegraphics[clip=true]{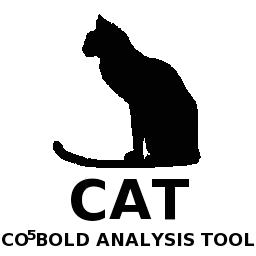}}
\caption{\footnotesize Logo of the \cobold\ Analysis Tool.} 
\label{fig:logo}
\end{figure}

The growing computational resources allow for increasingly larger and 
more detailed numerical simulations of stellar atmospheres, resulting 
in a considerably large amount of data. 
The production of advanced comprehensive models must therefore be 
accompanied by the development of efficient analysis and visualization 
software that is capable of handling the produced large data sets. 
Here, the \mbox{CO$^\mathrm{5}$BOLD} Analysis Tool (abbreviated CAT, 
Fig.~\ref{fig:logo}) is described. 
It is designed for an interactive analysis of 2D and 3D 
model atmospheres, which are produced with \cobold\ -- a   
widely used state-of-the-art code for the simulation of stellar atmospheres 
\citep{2012JCoPh.231..919F}.

\begin{figure*}[]
\centering
\resizebox{11.5cm}{!}{\includegraphics[clip=true]{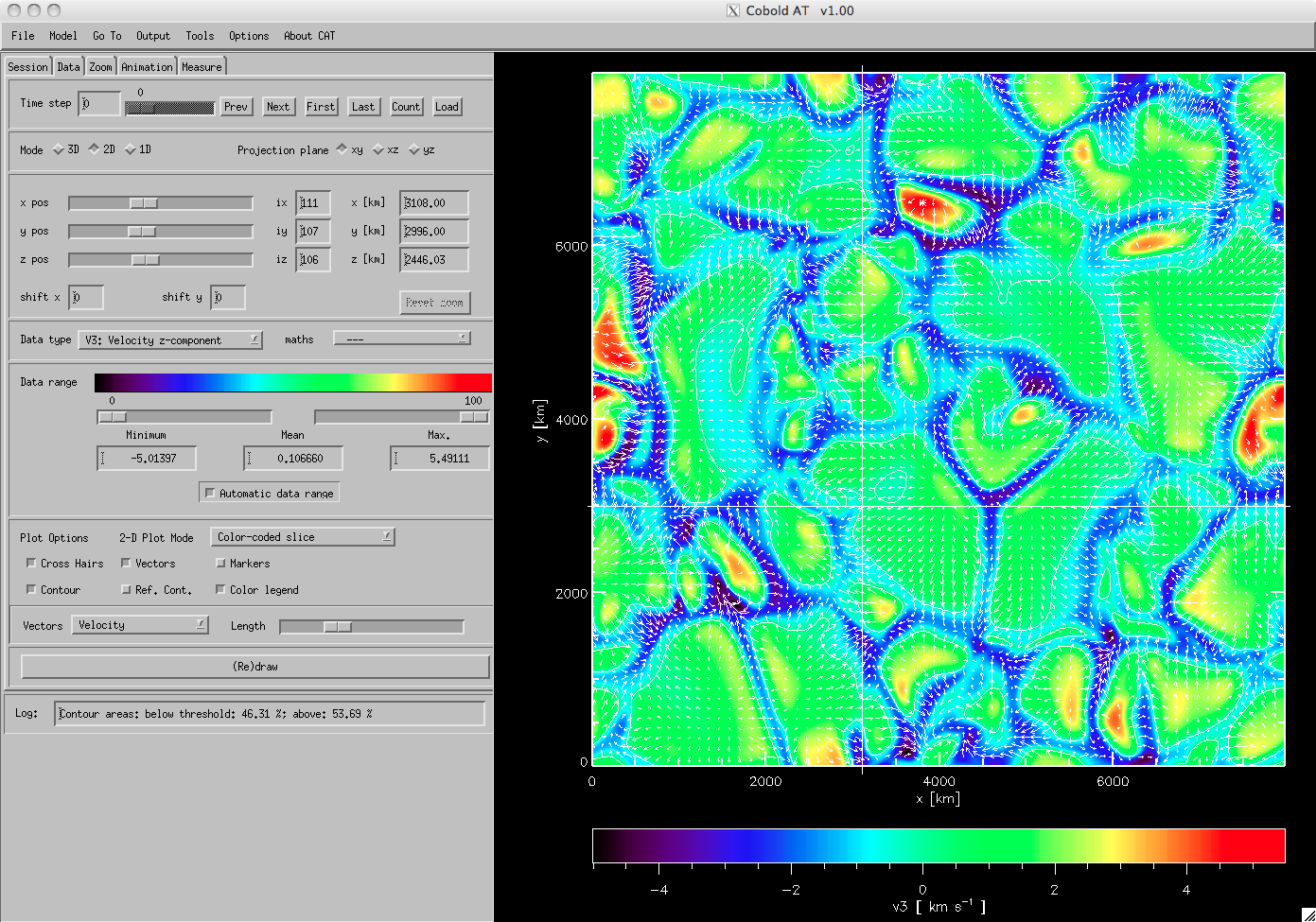}}
\caption{\footnotesize Screen shot showing the \textit{data tab} of the graphical user interface 
and the color-coded horizontal 2D slice through the photosphere of a 3D model of 
the solar atmosphere.
The selected physical quantity, here the vertical velocity, clearly shows 
the granulation pattern.  
The lines mark the current position of the interactive cursor.
 } 
\label{fig:screensht1}
\end{figure*}
\begin{figure*}[]
\centering
\resizebox{11.5cm}{!}{\includegraphics[clip=true]{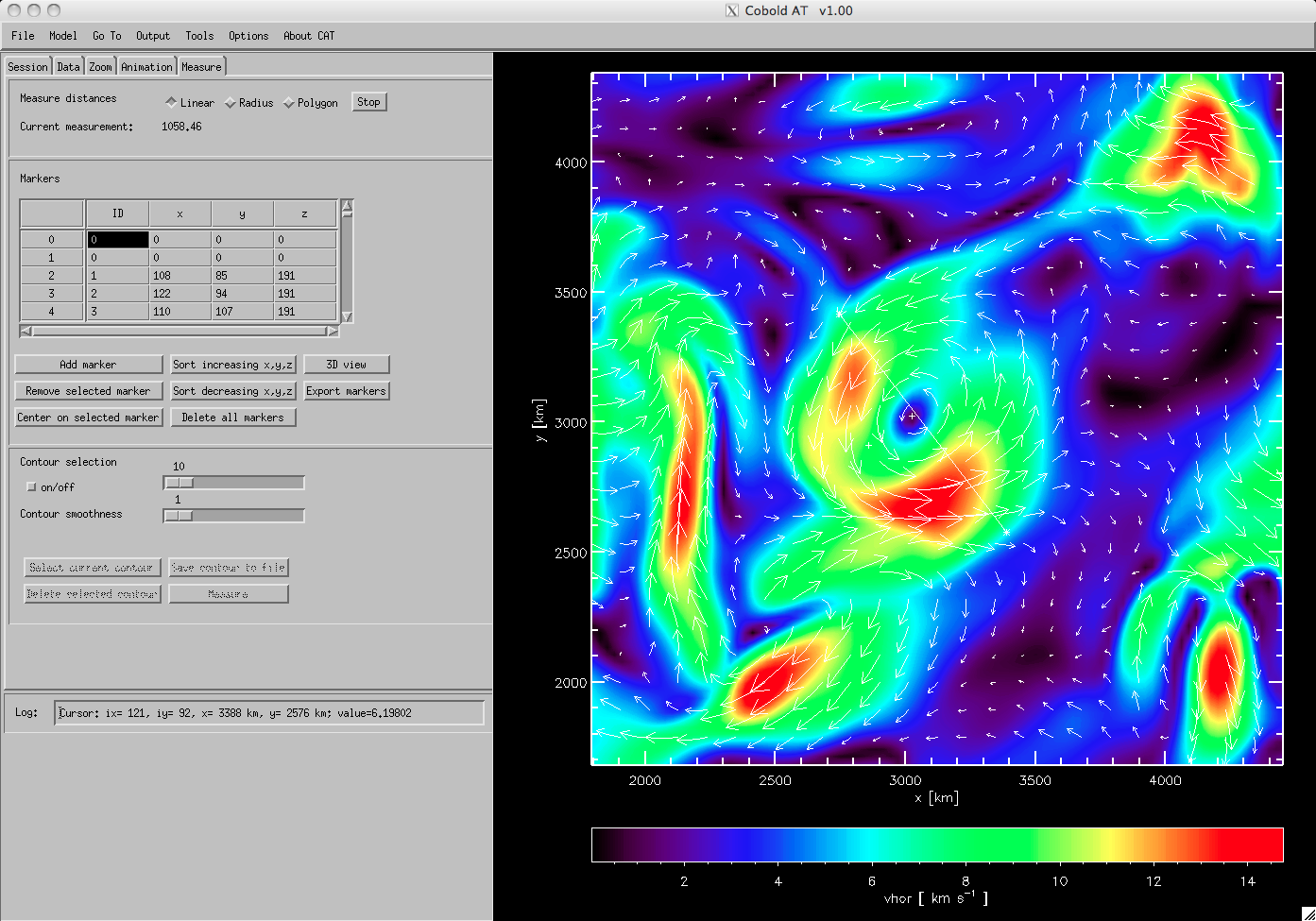}}
\caption{\footnotesize Screen shot showing the \textit{measure tab} 
of the graphical user interface and a 2D slice with the color-coded 
horizontal velocity. 
The displayed slice is a close-up region (zoom factor 3) in the chromosphere of the 3D MHD model 
of the solar atmosphere by \citet{2012Natur.486..505W}. 
The color shades and the arrows, which follow the horizontal 
velocity field, both exhibit a ring of increased velocity, which is caused 
by the rotation of the magnetic 
field structures.   
A diameter of 1058\,km is determined with the measurement tool 
(see upper left corner). 
} 
\label{fig:screensht2}
\end{figure*}

\section{Program overview}

CAT has an interactive graphical user interface, which is is programmed in IDL
(see Fig.~\ref{fig:screensht1}). 
The primary mode of operation is the display and analysis of slices through 
3D (or 2D) models. 
CAT can handle files with multiple snapshots (\textit{.full}). 

\paragraph{Settings and sessions.} 
After the installation, adjustments such as the default window size and 
paths can be made and saved. 
These standard settings will be restored every time CAT is started. 
Settings that are connected to a particular analysis session can be saved and restored 
as session files. 
It is possible to define a default session, which is automatically loaded when 
CAT starts.

\paragraph{Slice display}
The projection plane can be chosen perpendicular to the axes, 
resulting in the three planes \mbox{x-y}, \mbox{x-z}, and \mbox{y-z} in a 3D model.  
The position within a plane can be changed either by clicking in the displayed 
image or through the widgets in the control panel. 
CAT now also provides a zoom function in a separate tab, in which the 
magnification and the currently displayed region can be selected
(see Fig.~\ref{fig:screensht2} for a close-up region).

Different physical quantities can be chosen in the \textit{data tab} 
(see Fig.~\ref{fig:screensht1}), 
ranging from the basic quantities contained in the model file to 
more advanced quantities like, e.g., the spatial components of the electric current density   
\citep[e.g.,][]{2005ESASP.596E..65S}. 
The chosen quantity can then be combined with a mathematical operation, e.g. the logarithm, 
which is useful for quantities that cover many orders of magnitude in the displayed slice 
(e.g., the mass density in the \mbox{x-z} plane). 
By default, a color-coded 2D slice is displayed. 
Alternatively, the data in the current projection plane can be shown as surface plot, contour plot or line plot. 
Color-coded slices can be overlaid with contours (incl. a reference contour 
at optical depth unity) and/or a vector field. 
For the latter the spatial components of the velocity or the magnetic field in the 
selected projection plane can be drawn as vectors or streamlines, 
while the perpendicular component is not considered. 
The length of the vectors can be chosen directly in the \textit{data tab}
with more options being available in the \textit{options menu}.

\paragraph{Measurements.} 

The new \textit{measure tab} provides different tools for 
(i)~measuring distances between two points or along a polygon path, 
(ii)~defining markers, and 
(iii)~producing and exporting contour paths. 
The markers, which are sets of coordinates, and the contours can 
be interactively selected, modified and exported for further processing. 
These tools can be used for, e.g., defining 3D structures or for 
manually tracking features in space and time.  

\paragraph{Images and animation.} 
The displayed images can be output in various image formats. 
The animation tool provides an interactive way to produce MPEG videos 
or image sequences. 
It can either be a time animation for a series of model snapshots 
contained in one or more multi-snapshot file(s) or a 
spatial scan through a selected model snapshot along one of the 
spatial axes.

\paragraph{Export.} 
CAT offers several options to export model data. 
The currently loaded simulation snapshot can be written out in UIO format, 
even when it is part of a multi-snapshot (\textit{.full}) file. 
This feature is useful if a simulation terminated without producing an 
\textit{.end} file or if this file is corrupted for some reason. 
In this case CAT can be used to extract the last time step from the corresponding 
\textit{.full} file for using it as start model for the next simulation run. 
Model snapshots can also be saved as plane-parallel models. 
In a coming version, it will be possible to save only parts of the model 
and to extract data into IDL savefiles for further analysis.

\begin{figure}[]
\centering
\resizebox{\columnwidth}{!}{\includegraphics[clip=true]{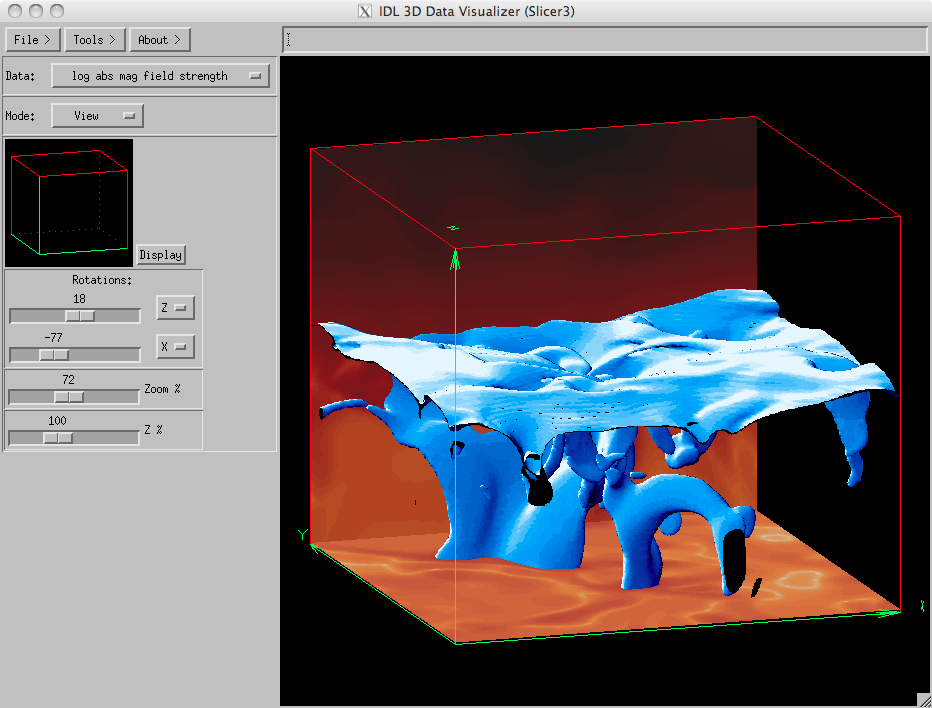}}
\caption{\footnotesize 
3D visualization of a MHD model of a M-type dwarf star 
\citep{2012arXiv1207.2342W}
with IDL as part of CAT. 
Plasma-beta is displayed as reddish slices and upper bluish surface. 
The lower blue surface enclose regions of high magnetic field strength. 
} 
\label{fig:screensht3}
\end{figure}

The 3D visualization tool, which is integrated in CAT, can give a first 3D 
impression but has limited functionality
(see Fig.~\ref{fig:screensht3}). 
The VAPOR tool \citep{vapor_clyne2007,vapor_clyne2005} allows for a  
more comprehensive analysis of the 3D structure, in particular for vector 
fields like velocity and magnetic field. 
For that reason, CAT offers the possibility to export \cobold\ data into 
VAPOR format.

\section{Data analysis examples}

CAT has been repeatedly proved useful in the past. 
It provides an easy way of monitoring the evolution of an ongoing simulation. 
CAT can also help to quickly reveal what is wrong  
in cases where a simulation terminates due to problems at a specific grid cell. 
Next to these ``care-taking'' tasks, CAT as been essential for the detailed scientific 
analysis in many cases. 

\paragraph{Swirls and tornadoes.} 
One example is the discovery of ring-like structures with increased horizontal 
velocity at chromospheric heights in 3D magnetohydrodynamic (MHD) models of the Sun.
Similar rings were previously observed as so-called ``chromospheric swirls'' 
with the Swedish 1-m Solar Telescope 
\citep{2009A&A...507L...9W}.
The rings of enhanced velocity, which appeared prominently in CAT
(see Fig.~\ref{fig:screensht2}), suggested that chromospheric swirls are the 
observational signature of rotating magnetic field structures. 
This finding finally led to a more comprehensive study of this phenomenon, then known 
as ``magnetic tornadoes'' \citep{2012Natur.486..505W}. 

\paragraph{M-type dwarf stars.}
CAT has been extensively used for the development of 
3D MHD atmosphere models of M-type dwarf stars that extend from the upper convection zone 
into the chromosphere \citep{2012arXiv1207.2342W}. 
The resulting set of models with different initial magnetic fields with 
different field strengths and different topologies 
(e.g., homogeneous vertical or mixed polarities) will be analysed in more 
detail in forthcoming publications.

\section{Outlook}

Since the first version in 2002, the development of CAT has been driven by the 
need for additional functionality for the analysis of increasingly complex 
models. 
Also in the future, CAT will be continuously extended. 
The feedback of users is therefore very welcome. 
Possible extensions in future releases could include the ability to read the data 
formats of other MHD codes and to display and analyse slices at arbitrary angles. 

\begin{acknowledgements}
The author likes to thank the organisers of the \mbox{2$^\mathrm{nd}$} \cobold\ Workshop 
(\mbox{CW$^\mathrm{2}$}), which was held in Heidelberg, Germany, in 2012. 
\end{acknowledgements}
\bibliographystyle{aa}
\bibliography{p13cat_sven} 
\end{document}